\documentclass[twocolumn,pre,showpacs]{revtex4}
\usepackage{graphicx}%
\usepackage{dcolumn}
\usepackage{amsmath}
\usepackage{latexsym}
\draft

\begin{document}
\def\bea{\begin{eqnarray}}
\def\eea{\end{eqnarray}}
\def\beq{\begin{equation}}
\def\eeq{\end{equation}}
\def\f{\frac}
\def\k{\kappa}
\def\sx{\sigma_{xx}}
\def\sy{\sigma_{yy}}
\def\sxy{\sigma_{xy}}
\def\e{\epsilon}
\def\ve{\varepsilon}
\def\ex{\epsilon_{xx}}
\def\ey{\epsilon_{yy}}
\def\exy{\epsilon_{xy}}
\def\be{\beta}
\def\D{\Delta}
\def\th{\theta}
\def\r{\rho}
\def\a{\alpha}
\def\s{\sigma}
\def\kb{k_B}
\def\la{\langle}
\def\ra{\rangle}
\def\nn{\nonumber}
\def\bu{{\bf u}}
\def\bn{\bar{n}}
\def\br{{\bf r}}
\def\up{\uparrow}
\def\dn{\downarrow}
\def\S{\Sigma}
\def\dg{\dagger}
\def\d{\delta}
\def\p{\partial}
\def\l{\lambda}
\def\G{\Gamma}
\def\o{\omega}
\def\g{\gamma}
\def\kv{\bar{k}}
\def\ha{\hat{A}}
\def\hv{\hat{V}}
\def\hg{\hat{g}}
\def\hG{\hat{G}}
\def\hTT{\hat{T}}
\def\h{\theta}
\def\uu{\vec{u}}
\def\rv{\vec{r}}
\def\hf{\frac{1}{2}}
\def\noi{\noindent}
\bibliographystyle{prsty}


\title{Heat conduction in a confined solid strip: Response to external strain} 
\author{Debasish Chaudhuri$^1$\footnote{debc@bose.res.in} and Abhishek Dhar$^2$\footnote{dabhi@rri.res.in}
}
\affiliation{
$^1$S. N. Bose National Centre for Basic Sciences,
Calcutta - 700098, India\\
$^2$Raman Research Institute,
Bangalore - 560080, India\\
}


\begin{abstract}
We study heat conduction in a system of hard disks confined to a
narrow two dimensional channel. The system is initially in a high density 
solid-like phase. We study, through nonequilibrium molecular dynamics 
simulations, the dependence of the heat current on an externally applied  
elongational strain. The strain leads to deformation and failure of the 
solid and we find that the changes in internal structure can lead to very
sharp changes in the heat current. A simple free-volume type
calculation of the heat current in a finite hard-disk system is proposed. 
This reproduces some qualitative features of the  current-strain graph
for small strains.
\end{abstract}
\maketitle

\section{Introduction}
\label{intro}
In a recent study \cite{myfail} it was observed that the properties of
a solid that is confined in a narrow channel can change drastically
for small changes in applied external strain. This was related to
structural changes at the microscopic level such as a change in the
number of layers of atoms in the confining direction. These effects
occur basically as a result of the small ( few atomic layers in one
direction ) dimensions of the system considered and confinement along
some direction. A similar layering
transition, in which the number of smectic layers in a confined liquid
changes in discrete steps with increase in the wall-to-wall separation, was
noted in \cite{degennes,landman}. 
Both \cite{myfail,degennes} look at  equilibrium properties while
\cite{landman} looks at changes in the dynamical properties. 
An interesting question is, how are transport properties, such as
electrical and thermal conductivity, affected for these nanoscale
systems under strain? This question is also important to address in
view of the  current interest in the properties of nanosystems, 
both from the point of view of fundamentals and applications 
\cite{datta,nanobook1,nanobook2}. 

In this paper we consider the effect of strain on the heat current
across a two-dimensional (2D) ``solid'' formed by a few layers of interacting
atoms confined in a long narrow channel.  
We note here that, 
in the thermodynamic limit it is expected that there can be no true
solid phase in this quasi-one-dimensional system. However for a long 
but finite channel, which is our interest here, and at a high
packing fraction the fluctuations are small and the system behaves 
like a solid. We will use the word ``solid'' in this sense.  

In Ref.~\cite{myfail} the anomalous failure, under strain, of
a narrow strip of  a 2D solid formed by hard disks confined within
hard walls [~see Fig.~\ref{cartoon}~] was studied. 
Sharp jumps in the stress vs strain plots were observed. These
were related to structural changes in the system which underwent  
transitions from solid-to-smectic-to-modulated liquid phases 
\cite{myfail,myijp}. 
In the present paper we  study changes in the thermal conductance of this 
system as it undergoes elastic deformation and failure
through a layering transition caused by external elongational
strains applied in different directions.

The calculation of heat conductivity in a many body system is 
a difficult problem. The Kubo formula and Boltzmann kinetic
theory provide formal expressions for the thermal conductivity. In
practice these are usually difficult to evaluate without making
drastic approximations. More importantly a large number of recent
studies \cite{bonet,lepri,lippi,grass} indicate that the heat conductivity of
low-dimensional systems infact diverge. It is then more sensible to
calculate directly the heat current or the conductance of the system rather than
the heat conductivity. In this paper
we propose a simple-minded calculation of the heat current which can be expected
to be good for a hard disk (or hard spheres in the three dimensional
case) system in the solid phase. This reproduces some qualitative
features of the simulations and gives  
values for the current which are of the correct order of magnitude.

The organization of the paper is as follows. In
Sec.~(\ref{simresults}) we explain the model and present the results
from simulations. In Sec.~(\ref{anaresults}) we derive a simple
formula for heat current in a hard-sphere system and evaluate it
approximately. We conclude with some discussions in Sec.~(\ref{disc}).  

\section{Results from simulations}

\label{simresults}
\begin{figure}[t]
\begin{center}
\includegraphics[width=7cm]{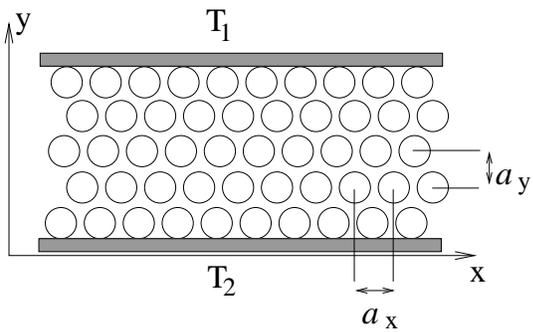}
\end{center}
\caption{ A solid with a triangular lattice structure formed by hard disks confined between two structureless walls at $y=0$ and $y=L_y$.
The walls are maintained at two different temperatures. The lattice
parameters of the unstrained solid are denoted by $a_x^0$ and $a_y^0$. 
Elongational strains can be imposed by rescaling distances either in
the $x$ or $y$ directions and the lattice parameters change to $a_x$
and $a_y$.
}
\vspace{1cm}
\label{cartoon}
\end{figure}
We consider a 2D system of hard disks of diameter $d$ and mass
$m$ which interact with each other through elastic collisions. 
The particles are confined  within a
narrow hard structureless channel [see Fig.~\ref{cartoon}]. 
The hard walls of the channel are located at
$y=0$ and $y=L_y$ and we take periodic boundary conditions in the $x-$direction.
The length of the channel along the $x-$direction is $L_x$ and the area is
${\cal A}=L_x\times L_y$.
The confining walls are maintained at  two 
different temperatures ( $T_2$ at $y=0$ and $T_1$ at $y=L_y$ ) so 
that the temperature difference $\D T=T_2-T_1$ gives rise to a heat 
current in the $y$-direction.
Initially we start with channel dimensions $L_x^0$ and $L_y^0$ such
that the system is in a phase corresponding to a unstrained solid with
a triangular lattice structure.
We then study the heat current in
this system when it is strained (a) along the $x-$direction and (b)
along the $y-$direction.

We perform an event-driven collision time dynamics \cite{allen} simulation of 
the hard disk system. The upper and lower walls are maintained at temperatures
$T_1 = 1$ and $ T_2 = 2$ (in arbitrary units) respectively by imposing Maxwell 
boundary condition \cite{bonet} at the two confining walls. This means
that whenever 
a hard disk collides with  either the lower or the upper wall it gets reflected
back into the system with a velocity chosen from the distribution
\bea
f(\bu)=\f{1}{\sqrt{2\pi}}\left(\f{m}{\kb T_W}\right)^{3/2}
|u_y|\exp\left(-\f{m\bu^2}{2\kb T_W}\right) 
\eea
where $T_W$ is the temperature ($T_1$ or $T_2$) of the wall on which
the collision occurs.  
During each collision energy is exchanged between the system and the
bath. Thus in our molecular dynamics simulation, the average heat
current flowing through the system can be found easily by computing the
net heat loss from the system to the two baths  (say $Q_1$
and $Q_2$ respectively) during a  large time interval $\tau$. 
The steady state heat current from lower to upper bath is given by $ \la I \ra
= \lim_{\tau \to \infty} Q_1/\tau = -\lim_{\tau \to \infty} Q_2 /\tau$.    
In the steady state the heat current (the heat flux density integrated over
$x$) is independent of $y$. This is a requirement coming from current
conservation. However if the system has inhomogeneities then the flux
density itself can have a spatial dependence and in general we can
have $j=j(x,y)$. In our simulations we have also looked at $j(x,0)$ and $j(x,L_y)$.

Note that the relevant scales in the problem are: $k_B T$ for energy, 
$d$ for length and $\tau_s=\sqrt{md^2/k_B T}$ for time. 
We start from a solid commensurate with its wall to wall separation and
follow two different straining protocols.
In case (a) we strain the solid by rescaling the length in the
$x$-direction and the imposed external strain is $\e_{xx} = (L_x -
L_x^0)/L_x^0$. In case (b) we rescale the length along the
$y$-direction and the imposed strain is $\e_{yy} = (L_y  
- L_y^0)/L_y^0$. 

\begin{figure}[t]
\begin{center}
\includegraphics[width=8.6cm]{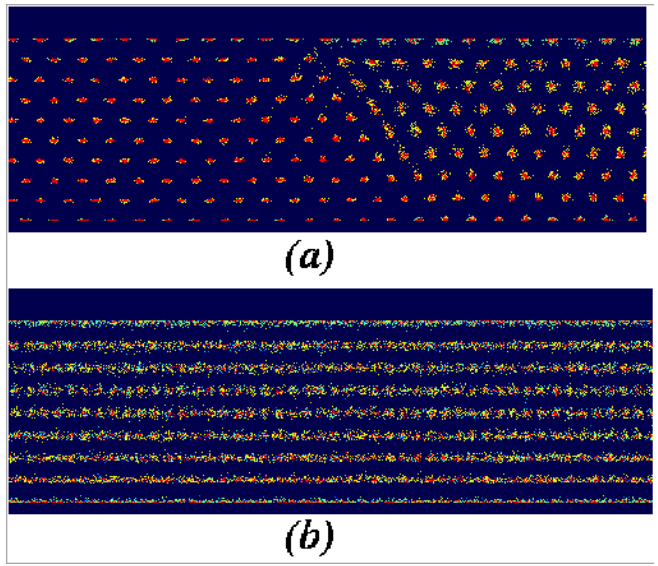}
\end{center}
\caption{
(Color online) Plots obtained by superposition of $500$ steady state
configurations of a portion of $40\times10$ system taken at equal 
time intervals. Starting from $\eta = 0.85$ imposition of strains 
(a) $\ex=0.1$, (b) $\ex=0.15$ gives rise to these structures.
The colors code local density of points from red/dark (high) to blue/light 
(low). In (a) one can see a $9$-layered structure  nucleated within a 
$10$-layered solid. The corresponding structure factor identifies this
 to be a smectic~\cite{myfail}. In
(b) the whole system has transformed into a $9$-layered smectic. 
}
\label{switch}
\end{figure}

\begin{figure}[t]
\begin{center}
\includegraphics[width=8.6cm]{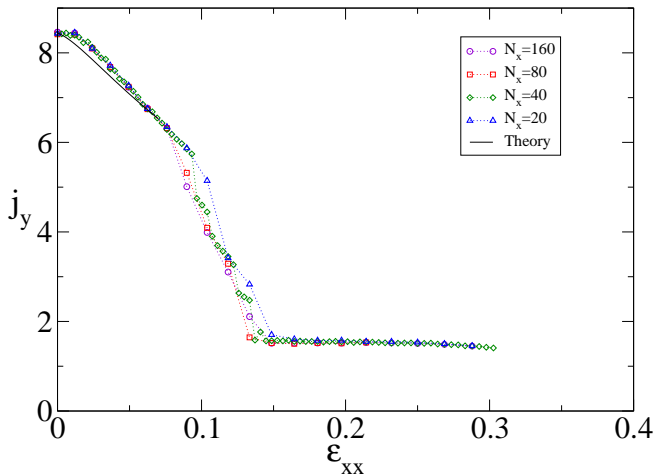}
\end{center}
\caption{ (Color online) Plot of $j_y$ vs $\ex$ 
   for different lengths of the channel. 
   In this graph as well as in all other view graphs we plot $j_y$
   in units of $\kb T/\tau_s d$.
   The width of the channel is
   $N_y=10$ layers. Starting packing fraction is $\eta = 0.85$. 
The solid line shows the theoretical prediction of
dependence of the heat current on strain [see Sec.~(\ref{anaresults})]. 
}
\label{jyex}
\end{figure}

The only thermodynamically  relevant variable for a hard disk system is the
packing fraction
$\eta = \pi N  d^2/4 {\cal A}$. For a close packed solid with periodic
boundary condition this value is about $\eta_c=0.9069$. On the other hand
for a confined solid having $N_y$ number of layers 
$\eta_c = \pi N_y/(2 \sqrt 3(N_y-1)+4)$ and for a $10$- layered solid
$\eta_c=0.893$.
In our simulations we consider initial
values of $\eta$ for the solid to be close to $\eta_c$. The channel
is ``mesoscopic'' in the sense that it has a small width with $N_y=10$
layers of disks in the $y-$direction (in the initially unstrained solid). In
the $x-$direction the system can be big and we consider
$N_x=20,~40,~80,~160$ number of disks in the $x-$direction.  
In collision time dynamics
we perform $10^5$ collisions per particle to reach the steady state and 
collect data over another $10^5$ collisions per particle.
All the currents calculated in this study are accurate within error bars which
are less than $3\%$ of the average current. 

Let us briefly mention some of the equilibrium results for the
stress-strain behavior obtained in Ref.~\cite{myfail}.
As the strain $\ex$ is imposed, the perfectly 
triangular solid shows rectangular distortion along with a linear response
in strain vs  stress behavior. Above a critical strain
($\ex \approx 0.1$) one finds that smectic bands having a lesser number
of layer nucleate within the solid [this can also be seen in Fig.~(2a) 
obtained from a nonequilibrium simulation]. 
This smectic is liquid-like in the $x$-direction (parallel to the
walls) and has  
solid-like density modulation order in the $y$-direction (perpendicular to the 
walls). With further increase in strain, the size of the smectic region 
increases and ultimately the whole system goes over to the smectic phase 
at $\ex \approx 0.15$ [Fig.~(2b)]. At even higher 
strains the smectic melts to a modulated liquid \cite{myfail,myijp}.
The corresponding structure factor shows typical liquid like ring 
pattern superimposed with smectic like density modulation peaks.
This layering transition is an effect of finite size in the confining
direction. Similar phase behavior has been observed in experiments on 
steel balls confined in quasi 1D \cite{pieranski-1}.
We note that, to fit a $N_y$ layered triangular solid within 
a channel of width $L_y$ we require
\bea
L_y = \frac{\sqrt 3}{2}a_x^0 (N_y-1) + d~. 
\eea
This enables us to define a fictitious number of layers 
\bea
\chi = 2 \frac{L_y- d}{\sqrt 3 a} +1 \nn
\eea
of triangular solid that can span the channel
where $a$ is the lattice parameter at any given density.
The actual number of layers that are present in the strained solid  is
$N_y = I(\chi)$ where the function $I(\chi)$ gives the  integer 
part of $\chi$. For confined solids the free energy has minima at
integer values of $\chi$ 
and maxima at half-integral values \cite{myijp,myfail}. The difference
in free-energy between successive 
maxima and minima gradually decreases with increasing 
$L_y$. Thereby the layering transition washes out for $n_l\gtrsim 25$ layered 
unstrained solid\cite{myfail}. Up to this number of layers, a triangular solid
strip confined between two planar walls fails at a critical deviatoric strain 
$\e^\ast_d\sim1/N_y$ (where $\e_d = \ex-\ey$). Smaller strips fail at
a larger deviatoric strain.

We now present the heat conduction simulation results for the two
cases of straining in $x$ and $y$ directions. 

{\bf{ (a)  Strain in $x$-direction}}~-~
In Fig.~\ref{jyex} we plot the heat current density $j_y$ calculated at
different values of the strain $\ex$. Starting from the 
triangular lattice configuration, we find that the heat current
decreases linearly with increase in strain. 
At about the critical strain $\ex \approx 0.1$ we find that the heat
current begins to fall at a faster rate.
This is easy to understand physically. At  the onset of critical
strain, smectic bands, which have lesser number of particle layers, start
nucleating (Fig.~\ref{switch}). These regions are much less effective 
in transmitting heat than the solid phase and the  heat current falls 
rapidly as the size of the smectic bands grow. 
At about the  strain value $\ex\approx 0.15$ the whole system is spanned by the
smectic.  Beyond this strain there is no appreciable change in the
heat current. 
The solid line in Fig.~\ref{jyex} is an estimate from a simple
analysis explained in Sec.~(\ref{anaresults}).

\begin{figure}[t]
\begin{center}
\includegraphics[width=8.6cm]{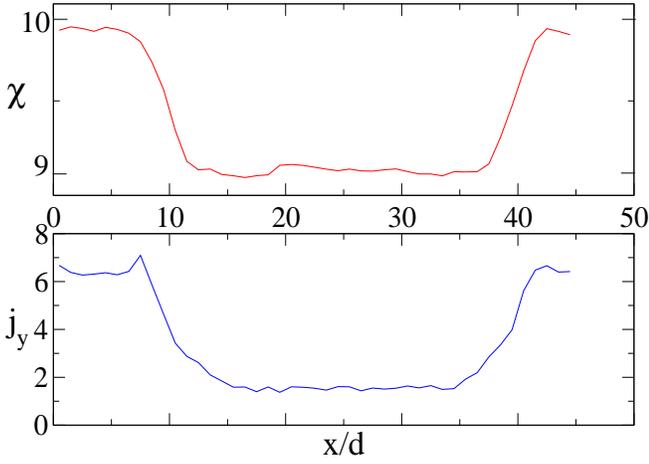}
\end{center}
\caption{ (Color online)
$\chi(x)$ is the local number of layers averaged over $10^3$
steady state configurations for a system of $40\times 10$ hard 
disks. A starting triangular lattice of $\eta=0.85$ is strained to $\ex=0.118$
and the data collected after steady state reached. Also shown is the local 
heat current $j_y(x)$. The regions having lower number of layers conduct 
less effectively.
}
\label{jy-x}
\end{figure}

In Fig.~\ref{jy-x} we plot the local steady state heat current
$j_y(x)$ for a system of $40\times 10$ particles  at a 
strain $\ex=0.118$ {\emph{i.e.}} at a strain corresponding to the
solid-smectic phase 
coexistence. At this same strain the number of layers averaged over $10^3$
configurations have been plotted. It clearly shows that 
the local heat current is smaller in regions with smaller number of
layers. This is the reason behind getting a sharp drop in average heat
current after the onset of phase coexistence.

\begin{figure}[t]
\begin{center}
\includegraphics[width=8.6cm]{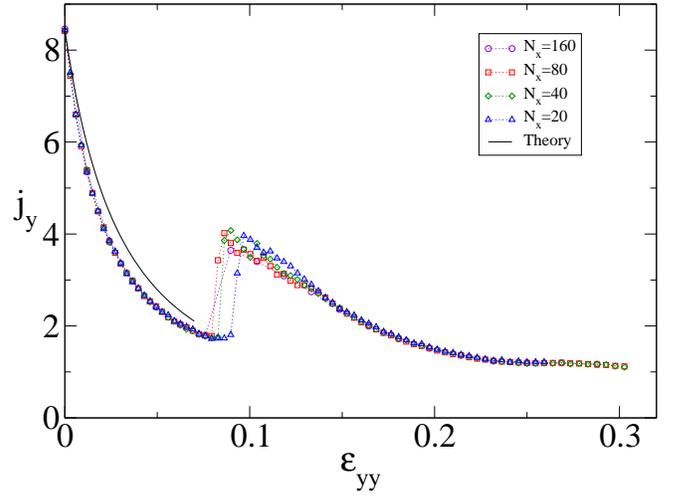}
\end{center} 
\caption{ (Color online) Plot of $j_y$  vs $\ey$ for different channel
  lengths. The channel width is $N_y=10$ layers. The starting packing
  fraction is $\eta =0.85$. The jump in current occurs at the strain
  value where the number of layers in the $y-$direction increases by
  one and the system goes to a   smectic phase. 
The solid line shows the theoretical prediction of
dependence of the heat current on strain [~see
  Sec.~(\ref{anaresults})~].
\vspace{1cm}
}
\label{jyey}
\end{figure}

\begin{figure}[t]
\begin{center}
\includegraphics[width=8.6cm]{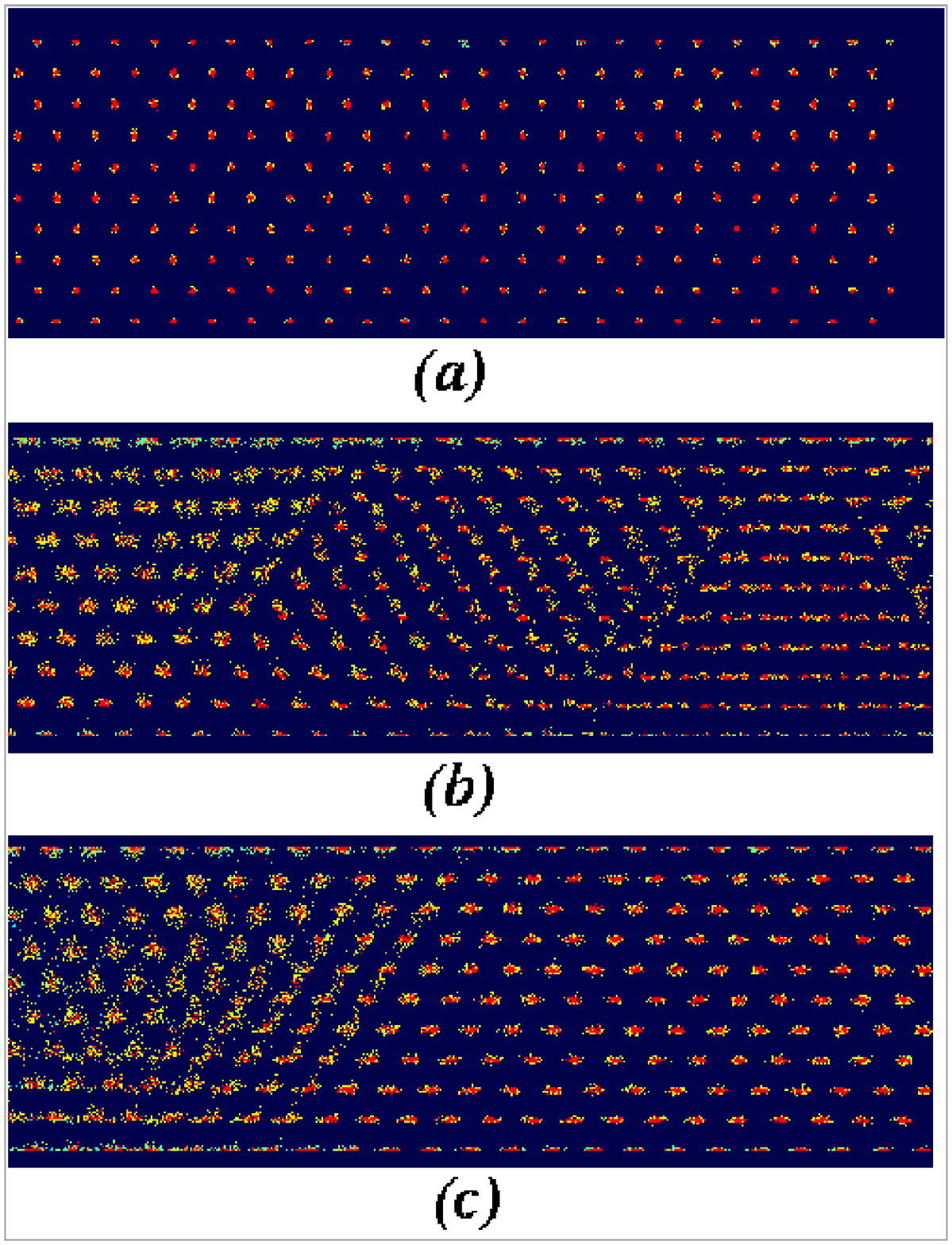}
\end{center}
\caption{
(Color online) Plots obtained by superposition of $500$ 
steady state configurations of a portion of $40\times10$ system taken at equal 
time intervals. Starting from $\eta = 0.85$ imposition of strains 
(a) $\ey=0.05$, (b) $\ey=0.1$, (c) $\ey=0.12$ gives rise to these structures.
The colors code local density of points from red/dark (high) to blue/light
(low). (a) Solid phase. (b) A mixture of $10$-layered solid and a buckling 
phase. (c) An $11$-layered solid in contact with $10$-layered smectic like
region. 
}
\label{jump}
\end{figure}

{\bf{(b)Strain in $y$-direction}}~-~ Next we consider the case where,
again starting from the density  
$\eta =0.85$, we impose a strain along the $y-$direction. As shown in 
Fig.~\ref{jyey}, the heat current $j_y$ now has  
a completely different nature. The initial fall is much steeper and 
has a form different from the linear drop in Fig.~\ref{jyex}. 
The approximate analytic curve is explained in Sec.~(\ref{anaresults}).  
At about $\ey\approx 0.1$ we see a sharp and presumably discontinuous
jump in the current. At this point the system goes over to a buckled phase
(Fig.~\ref{jump}b) in which different parts of the solid (along $x$-direction)
are displaced along the $y$-direction by small amounts so that the
extra space between 
the walls is covered \cite{buckled-1,buckled-2,buckled-3}. A further
small strain induces a layering transition and the system breaks into
two regions one of which is an $N_y=11$ layered solid and the other is
a $N_y=10$ layered highly fluctuating smectic-like region. At even higher 
strains ($\ey\sim 0.2$) the whole system eventually melts to an $N_y=11$ 
layered smectic phase. The phase behavior of this system is interesting 
and will be discussed in detail elsewhere\cite{myfail-large}.
Unlike the case where the applied strain is in the $x$-direction,
in the present case the buckling-layering transition is very
sharp. Even though the overall density has decreased, due to
buckling and increase in number of layers in the conducting direction,
there is an increase in the energy transferring collisions and hence
the heat current.  
The plots in Fig.~\ref{jump} show the structural
changes that occur in the system as one goes through the transition.

\begin{figure}[t]
\begin{center}
\includegraphics[width=8.6cm]{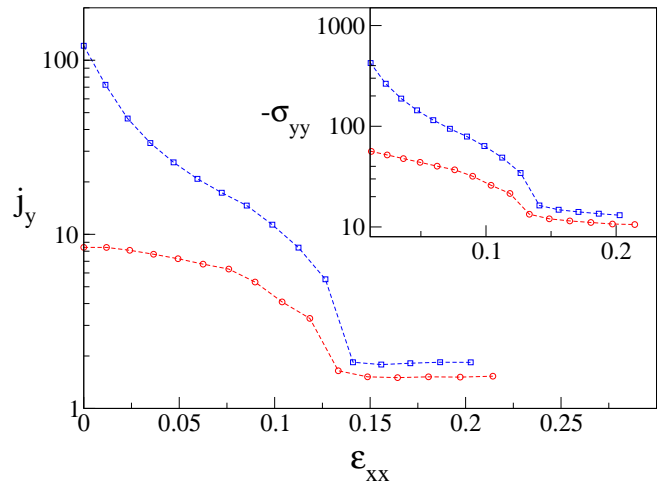}
\end{center}
\caption{(Color online) Plot of $j_y$   vs $\ex$
  for two different starting values of 
  the packing fraction. $\Diamond$ corresponds to a starting value of 
$\eta = 0.89$ while  $+$ is for $\eta = 0.85$. 
In both the cases the initial solid size was $80 \times 10$. The inset shows 
corresponding plots of $-\sy$ (in units of $k_B T/ d^2$) vs
  $\ex$. Notice that stress-strain curve has the 
same qualitative profile as the $j_y$ vs $\ex$ curve.
}
\label{jyex-comp}
\end{figure}

We find in general that the heat current along any direction within the solid
shows the same qualitative features as the stress component along the same
direction. 
This can be seen in Fig.\ref{jyex-comp} where we have plotted $j_y$
vs $\ex$ for two starting  densities of solids $\eta =
0.85,~0.89$. In the inset we show the corresponding  $-\s_{yy}$ vs $\ex$
curves and see that they follow the same qualitative  
behavior as the heat current curves. 
The reason for this is that  microscopically
they both originate from interparticle collisions.
Infact the microscopic expressions for the total heat current 
[~see Eq.~(\ref{totI}) in Sec.~(\ref{anaresults})~] is very 
similar to that for the stress tensor component, with an extra velocity factor. The stress tensor is given by:
\bea
 {\cal A}\s_{\a \be} &=&     -\sum_{i} \la m u_i^\a u_i^\be \ra 
+\sum_{i< j
  }  \left\la \f{\p \phi(r_{ij})}{\p   r_{ij}} \f{x_{ij}^\a
  x_{ij}^\be}{r_{ij}}  \right\ra ~,
\label{sts}
\eea
where $\{x_i^\a, u_i^\a \}$ refer to the $\a$-th component of 
position and velocity of the $i^{\rm th}$ particle, 
$r_{ij}^2=\sum_\a (x_{ij}^\a)^2$ and $\phi(r_{ij})$ is the interparticle 
potential. For a hard disk system, $\f{\p\phi(r_{ij})}{\p
  r_{ij}}$ 
can be replaced by $-\kb T\d(r_{ij}-{\rm d})$. Also 
in equilibrium we have $\la m u_i^\a u_i^\be \ra = \kb T \d_{\a\be}$
and hence the stress tensor becomes:
\bea
{\cal A} \s_{\a \be} =  -\kb T \left[  N  \d_{\a\be}+  \left\la \sum_{i<j}  
   \f{x_{ij}^\a x_{ij}^\be}{r_{ij}} \d(r_{ij}(t)- d)\right\ra 
~\right] \nn~.
\eea
Using collision time simulation it is easier to evaluate the 
stress tensor in the following way. 
We can rewrite Eq.~(\ref{sts}) as
\bea
{\cal A} \s_{\a \be} &=&     -N \kb T \d_{\a\be} -\sum_{i<j}  
     \left\la  x_{ij}^\a f_{ij}^\be \right\ra \nn~.
\eea 
We use the fact that $\la\dots\ra$ can be
replaced by a time average so that from Eq.~(\ref{sts}) we have
\bea
\ \left\la  x_{ij}^\a f_{ij}^\be \right\ra =-\lim_{\tau \to \infty} \f{1}{\tau} \int_0^\tau dt 
 x_{ij}^\a f_{ij}^\be \nn~.
\eea
Now note that during a collision we have 
$\int dt f_{ij}^\be= \Delta p_{ij}^\be$
where $\Delta \vec p_{ij}$ 
is the  
change in momentum of $i^{\rm th}$ particle due to collision with
$j^{\rm th}$ particle. 
It can be shown that $\Delta {\bf{p}}_{ij} = -(\bu_{ij}.\hat{ \br}_{ij})\hat
{\br}_{ij}$  where  $ \hat{ \br}_{ij} = \br_{ij}/r_{ij}$ and $\bu_{ij}=
\bu_i-{\bu}_j$ and $\br_i,~\bu_i$ are evaluated just before a
collision.  
This change in momentum occurs for a single pair of particle during one
collision event. 
To get the stress tensor we sum over all the 
collision events in the time interval $\tau$ between all pairs of
particles. Therefore for collision time dynamics  
we get the following expression for the stress tensor,
\bea
{\cal A} \sigma_{\a\be} =  -N \kb T \d_{\a\be}+
\lim_{\tau \to \infty} \f{1}{\tau} \sum_{\tau_c} \sum_{i<j} \Delta
p_{ij}^\a x_{ij}^\be ~,
\eea
where $\sum_{\tau_c}$ denotes a summation over all collisions in time
$\tau$. 

\section{Analysis of qualitative features}
\label{anaresults} 
We briefly outline a derivation of the expression for the heat
flux. For the special case of a hard disk system this simplifies
somewhat. We will show  
that starting from this expression and making rather simple minded
approximations  we can explain some of the observed results for  heat flux as a function 
of imposed external strain.

We consider a system with a general Hamiltonian given by:
\bea
H=\sum_i [\f{m \bu_i^2}{2}+V(\br_i)] +\f{1}{2}\sum_{i,j \neq i} \phi(r_{ij})~, 
\eea
where $V(\br_i)$ is an onsite potential which also includes the wall. 
To define the heat current density we need to write a continuity
equation of the form: $\p \e(\br,t)/\p t+\p j_\a(\br,t)/\p x_\a=0$. 
The local energy density is given by: 
\bea
\e(\br,t)&=&\sum_i \d (\br-\br_i) h_i {\rm~~~where} \nn \\ 
h_i &=& \f{m \bu_i^2}{2}+V(\br_i) +\f{1}{2} \sum_{j \neq i} \phi(r_{ij}) \nn
\eea
Taking a derivative with respect to time gives
\bea
\f{\p \e}{\p t} &=& -\f{\p}{\p x_\a} \sum_i \d (\br - \br_i) h_i u_i^\a+
\sum_i \d (\br-\br_i) \dot{h}_i   \\
&=& -\f{\p}{\p x^\a} j_{\a}^K + W^U 
\eea
where ${\bf j}^K=\sum_i \d (\br - \br_i) h_i \bu_i $ is the convective
part of the energy current. We will 
now try to write the remaining part given by $W^U$ as a divergence
term. We have
\bea
W^U &=& \sum_i \d (\br-\br_i) \dot{h}_i \nn \\
&=& \sum_i  \d (\br-\br_i) [m u_i^\a \dot{u}_i^\a + \f{\p V(\br_i)}{\p
    x_i^\a} u_i^\a \nn\\
 &-& \f{1}{2} \sum_{j \neq i} (f_{ij}^\a  u_i^\a + f_{ji}^\a u_j^\a)
]~, \nn
\eea
where $f_{ij}^\a=-\p\phi(r_{ij})/\p x_i^\a$.
Using the equation of motion $m \dot{u}_i^\a = -{\p V}/{\p
  x_i^\a}+\sum_{j\neq i} f_{ij}^\a$ 
we get 
\bea
W^U &=&  \f{1}{2}  \sum_{i,j \neq i} \d (\br-\br_i) (~f_{ij}^\a
u_i^\a - f_{ji}^\a u_j^\a~)~.
\eea
With the identification $W^U=-{\p j_\a^U}/{\p x^\a}$ and using
${\bf{f}}_{ij}=-{\bf{f}}_{ji}$ we finally get:
\bea
j_\a^U(\br) 
= \f{1}{2} \sum_{i,j \neq i} \theta (x_i^\a-x^\a)
\prod_{\nu  \neq \a} \d (x^\nu-x_i^\nu)f_{ij}^\be( u_i^\be + u_j^\be)~   
\eea
where $\theta(x)$ is the Heaviside step function.
This formula has a simple physical
interpretation. First note that we need to sum over only those $i$ for
which $x_i^\a > x^\a$. Then the formula basically gives us the net rate at
which work is done by particles on the left of $x^\a$ on the particles
on the right which is thus the rate at which energy flows from left to
right. The other part, $j_\a^K$, gives the energy flow as a result of
physical motion of particles across $x^\a$.  
Let us look at the total current in the system. Integrating the
current density $j^U_\a$ over all space we get:
\bea
I_\a^U &=& \f{1}{2} \sum_{i,j\neq i} x_i^\a f_{ij}^\be (u_i^\be + u_j^\be) \nn\\
&=& - \f{1}{2} \sum_{i,j\neq i} x_i^\a \f{\p \phi (r_{ij})}{\p r_{ij}}
\f{x^{\be}_{ij}}{r_{ij}} (u_i^\be + u_j^\be) \nn \\
&=& - \f{1}{4} \sum_{i,j\neq i}  \f{\p \phi (r_{ij})}{\p r_{ij}}
\f{x^\a_{ij} x^{\be}_{ij}}{r_{ij}} (u_i^\be + u_j^\be)  ~.
\eea
Including the convective part and taking an average over the steady
state we finally get:
\bea
\la I_\a \ra &=& \la I_\a^K \ra + \la I_\a^U \ra = 
 \sum_i \la~ h_i u_i^\a \ra \nn \\
&-&\f{1}{4} \sum_{i ,j \neq
  i} ~ \la~ \f{\p \phi (r_{ij})}{\p r_{ij}}
\f{x^\a_{ij} x^{\be}_{ij}}{r_{ij}} (u_i^\be + u_j^\be)~ \ra~. 
\label{totI}
\eea
We note that for a general phase space variable $A(\{x_i,u_i\})$ the average $\la
A \ra$ is the time average $\lim_{\tau \to \infty} (1/\tau)
\int_0^\tau dt A(\{x_i(t),u_i(t)\})$.

{\bf Finding the energy current for a hard disk system}: 
The energy current expression
involves the velocities of the colliding particles which change during
a collision so we have to be careful. 
 We use the following expression
for $\la I_\a^U \ra$:
\bea
\la~I_\a^U~\ra &=& \f{1}{4} \sum_{i, j \neq i} \la~ x_{ij}^\a 
({f_{ij}}^{\be} u_i^\be 
-f_{ji}^\be  u_j^\be)~ \ra \nn \\ &=& \lim_{\tau \to \infty} \f{1}{\tau}
\int_0^\tau dt \f{1}{4} \sum_{i, j\neq i}  x_{ij}^\a (f_{ij}^\be
u_i^\be -f_{ji}^\be  u_j^\be)
\eea
Now if we integrate across a collision we see that $\int dt ({\bf
  f}_{ij}.\bu_i)$ gives the change in kinetic energy of the $i^{\rm{th}}$
particle during the collision while $ \int dt ({\bf f}_{ji}.\bu_j)$
gives the change in kinetic energy of the $j^{\rm{th}}$ particle. Hence we get
\bea
\la~I_\a^U~\ra &=& \sum_{i, j \neq i} \lim_{\tau \to \infty} \f{1}{\tau}  \sum_{t_c} \f{1}{4}
x_{ij}^\a (\Delta K_i - \Delta K_j) \nn \\ &=& \sum_{i < j}  \f{\la~ 
x_{ij}^\a \Delta K_i~ \ra_c}{\la~\tau_{ij}~\ra_c}
\label{curr}
\eea
where we have used the fact that for elastic collisions $\Delta K_i =
-\Delta K_j $ and $\sum_{t_c}$ denotes a summation over all
collisions, in the time interval $\tau$, 
between pairs $\{ij\}$. The time interval between successive
collisions between $i^{\rm th}$ and $j^{\rm th}$ particles is
denoted by $\tau_{ij}$ and the average $\la~...~\ra_c$ in the last line  denotes a
{\emph{collisional}} average. Thus $\la \tau_{ij} \ra_c =\lim_{\tau \to
    \infty} \tau/N_{ij}(\tau)$, where $N_{ij}(\tau)$ is the number of
  collisions between $i^{\rm th}$ and $j^{\rm th}$ particles in time $\tau$.  
For hard spheres the convective part of the current involves only the
kinetic energy and is given by
$\la ~I_\a^K~\ra =\sum_i \left\la~ (m\bu_i^2 /2) u_i^\a~ \right\ra$.
Using these expressions we  now try to obtain estimates of the heat
current and its dependence on strain (near the close packed limit where
the system looks like a solid with the structure of a strained
triangular lattice). 

Near the close packed limit the convection current can be neglected and
we focus only on the conductive part given by $\la I^U \ra = \la I^U_2
\ra$ (for conduction along the $y-$direction).
At this point we assume local thermal equilibrium (LTE) which we prove
from our simulation data at the end of this section.
Assuming LTE we write the following approximate
form for the energy change $\Delta K_i$ during a collision:
\bea
\D K_{i} =  \kb (T(y_j)-T(y_i)) = -\kb \f{dT}{dy} y_{ij}=y_{ij}
    \f{\kb \D T}{L_y}~, \nn
\eea
where we have denoted $x^{(\a=2)}_i=y_i$ and $\D T=T_2-T_1$. The
temperature gradient has been assumed  to be 
small and constant. Further we assume that in the close packed limit
that we are considering, only nearest neighbor pairs $\{<ij>\}$ contribute
to the current in Eq.~(\ref{curr}) and that they contribute equally. 
We then get the following approximate form for the total current:
\bea
{\la I_y \ra } \approx \f{3N\kb \D T}{L_y} \f{  y_c^2}{\tau_c}~, 
\label{iu}
\eea
where $\tau_c$ is the average time between successive collisions
between two particles while $y_c^2$ is the mean square separation along 
the $y-$axis of the colliding particles. 
Finally, denoting the density of particles by $\r=N/{\cal A}$ we get
for the current density: 
\beq
j_y = \f{\la I_y \ra}{\cal A} \approx \f{3 \r \kb \D T }{  L_y} \f{  y_c^2}{\tau_c}~. 
\label{ju}
\eeq

\begin{figure}[t]
\begin{center}
\includegraphics[width=7.0cm]{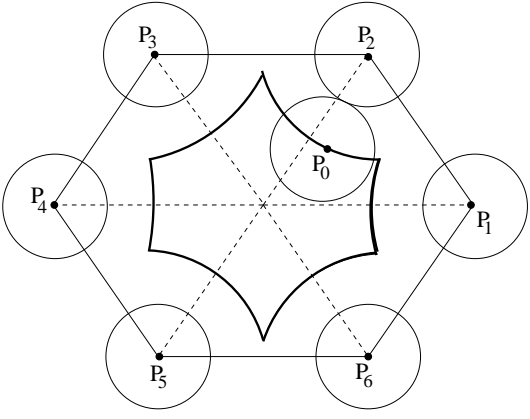}
\end{center}
\caption{In our free-volume theory we assume that the outer six disks
  are fixed 
and the central disk moves within this cage of fixed particles. 
The curve in bold line shows the boundary $\cal{B}$ of the free
  volume. A point on this boundary is denoted by $P_0(x,y)$ while the
  centers of the six fixed disks are denoted by $P_i(x_i,y_i)$ with
  $i=1,2...6$. 
}
\label{fvol}
\end{figure}

For strains $\ex$ and $\ey$ in the $x$ and $y$ directions we have
$\r=\r_0/[(1+\ex)~(1+\ey)]$. We estimate $y_c^2$ and $\tau_c$ from a
simple equilibrium free-volume theory, known as fixed neighbour free
volume theory (FNFVT). In this picture we think
of a single disk moving in a fixed cage formed by taking the average
positions of its six nearest neighbor disks [see Fig.~\ref{fvol}]. For
different values of the strains we then evaluate the average values 
$[y_c^2]_{fv}$ and $[\tau_c]_{fv}$ for the moving particle from FNFVT.
\begin{figure}[t]
\begin{center}
\includegraphics[width=6.5cm]{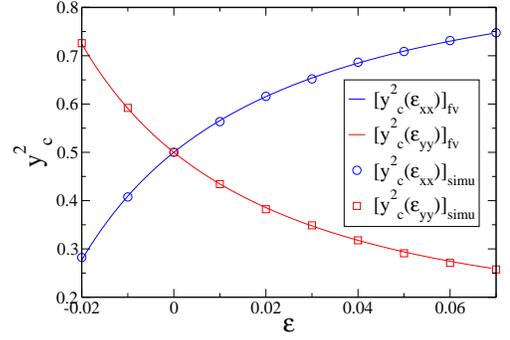}
\vskip 1.2cm
\includegraphics[width=6.5cm]{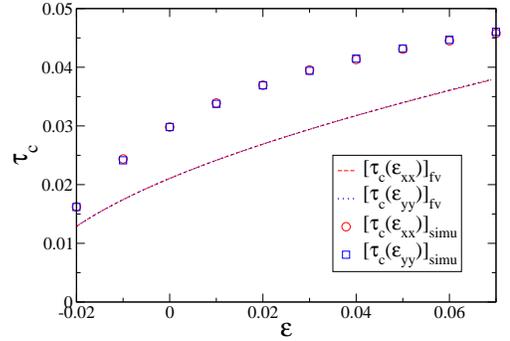}
\end{center}
\caption{ (Color online)
  Plots showing comparison of the analytically calculated
  values of $[y_c^2]_{fv}$ (in units of $d^2$)
  and $[\tau_c]_{fv}$ (in units of $\tau_s$) with those obtained
  from a free volume simulation of a single disk moving within the
  free volume cage. The free volume corresponds to a
  starting unstrained triangular lattice at $\eta=0.85$ which is then
  strained along $x$ or $y$ directions.  
}
\label{fv-y2tc}
\end{figure}
We assume that the position of the center of the moving disk
$P_0(x,y)$, at the
time of collision with any one of the six fixed disks, is uniformly
distributed  on the boundary $\cal{B}$ of the free-volume. Hence
$[y_c^2]_{fv}$ is easily calculated using the expression:
\bea
[y_c^2]_{fv}=\f{\sum_i  \int_{{\cal{B}}_i} ds
  (y-y_i)^2}{L_{\cal{B}}}~,\label{ycfv} 
\eea 
where ${\cal{B}}_i$ is the part of the boundary $\cal{B}$ of the free
volume when the middle disk is in contact with the $i^{\rm{th}}$ fixed
disk, $ds$ is the infinitesimal length element on $\cal{B}$ while
$L_{\cal{B}}$ is the total length of $\cal{B}$.  
Let the unstrained lattice parameters be
$a_x^0,~a_y^0=\sqrt{3}a_x^0/2$. Under 
strain we have $a_x=a_x^0(1+\e_{xx})$ and $a_y=a_y^0 (1+\e_{yy})$.  
Using elementary geometry we can then evaluate $[y_c^2]_{fv}$ from
Eq.~(\ref{ycfv}) in terms of $\e_{xx},~\e_{yy}$ and the unstrained
lattice parameter $a_x^0$. An exact calculation of $[\tau_c]_{fv}$ is
nontrivial.  
However we expect $[\tau_c]_{fv}=c~V_{fv}^{1/2}/T^{1/2} $ where
$V_{fv}$ is the ``free 
volume'' [see Fig.~\ref{fvol}] and   $c$ is a constant factor of
$O(1)$ which we will use as a fitting parameter. The calculated values
for $[y_c^2]_{fv}$ and $[\tau_c]_{fv}$ (see the appendix) are shown in  Fig.~\ref{fv-y2tc}. 
Also shown are their values obtained from an equilibrium simulation of
a single disk moving inside the free volume cage.   
Thus we obtain the following estimate for the heat current:
\bea
[j_y]_{fv}= \f{3 \r \kb T^{1/2} \D T }{  L_y} \f{[y_c^2]_{fv} }{c~V_{fv}^{1/2}}~. 
\eea
We plot in Fig.~(\ref{jyex}) and Fig.~(\ref{jyey}) the above estimate of $[j_y]_{fv}$
along with the results from simulations. We find that the overall features 
of the simulation are reproduced with $c=0.42$. 

\begin{figure}[t]
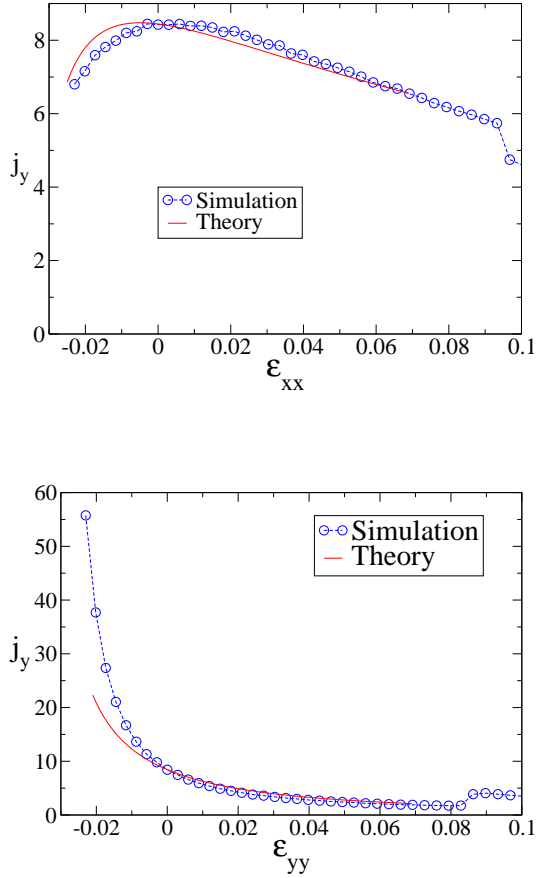

\begin{center}
\includegraphics[width=7.0cm]{jy-ex-negpos.eps}
\vskip 1.2cm
\includegraphics[width=7.0cm]{jy-ey-negpos.eps}
\end{center}
\caption{ (Color online)
  Plot showing effect on $j_y$ of negative strains applied in
  the $x$ and $y$ directions. 
The system is prepared initially in  a triangular lattice at
  $\eta=0.85$. Note that negative $\ex$ reduces $j_y$ whereas negative
  $\ey$ increases $j_y$.
}
\label{comprs}
\end{figure}
\begin{figure}[t]
\begin{center}
\includegraphics[width=8.6cm]{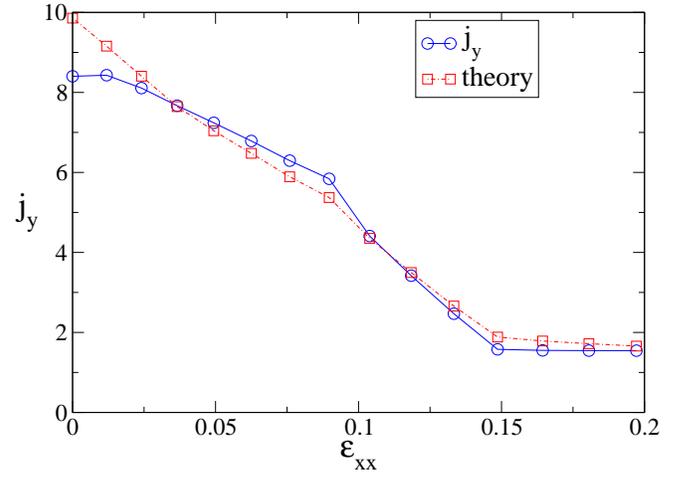}
\end{center}
\caption{(Color online)
  Comparison of simulation results for $j_y$ with the
  approximate formula   in Eq.~(\ref{ju}) where   
$\tau_c$ and  $ y_c^2$ are also calculated directly from the same
  simulation. The results are for a $40\times10$ system with starting
  value of   $\eta=0.85$ and strained along $x-$direction. 
}
\label{simu-y2tc}
\vspace{1cm} 
\end{figure}
For small strains we find (see the appendix) $[y_c^2(\ex)]_{fv} \sim
0.5+\a \ex - \be_1\ex^2 $,~ 
$[y_c^2(\ey)]_{fv}\sim0.5-\a \ey+ \be_2\ey^2 $ and
$[\tau_c]_{fv} \sim(\g_1 +\g_2 \e - \g_3 \e^2)$ where $\e$ stands for 
either $\ex$ or $\ey$ and  $\a,~\be_1,~\be_2,~\g_1,\g_2,\g_3$ are all 
positive constants that depend only on
$a_x^0$.
For $\eta=0.85$ we have  $\a=7.62,~\be_1=121.77,~\be_2=124.37$ and 
$\g_1=0.02$, $\g_2=0.33$, $\g_3=1.125$.
From these small strain scaling forms we find that $j_y(\ey)$
always decreases with positive 
$\ey$ and increases with negative or compressive $\ey$ 
(note that we always consider starting configurations of a triangular solid
of any density). 
On the other hand the sign of the change in $j_y(\ex)$ will
depend on the relative magnitudes of $\a,~ \be_1$ and $\g$ . For
starting density 
$\eta=0.85$, $j_y(\ex)$ decreases both for positive and negative $\ex$. 
In Fig.~\ref{comprs} we show the effect of compressive strains $\ex$ and 
$\ey$ on the heat current $j_y$ and compare the simulation results
with the free volume theory. 

It is possible to calculate $y_c^2$ and $\tau_c$ directly  from our  
nonequilibrium collision time dynamics simulation. The mean collision
time $\tau_c$ is obtained by dividing the total simulation time
by the total number of collisions per colliding pair.
Similarly $y_c^2$ is evaluated at every collision and we then obtain
its average. Inserting these values of $\tau_c$  and $\la y_c^2\ra$  into
the right hand side of Eq.~(\ref{ju}) we get an estimate of the
current as given by our theory (without making use of free-volume
theory). In Fig.~\ref{simu-y2tc} we compare this value of the current
$j_y$, for strain $\e=\ex$, and compare it with the simulation
results. The excellent agreement between the two indicates that our simple
theory is quite accurate.

\begin{figure}[t]
\begin{center}
\includegraphics[width=7.0cm]{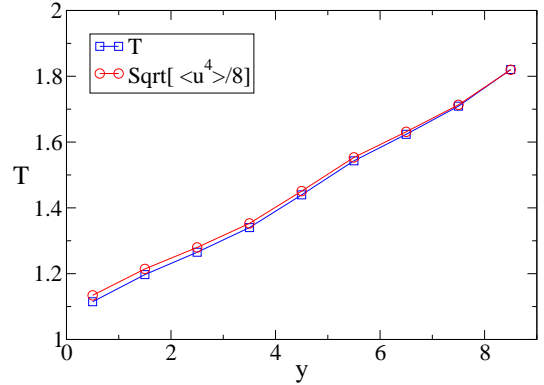}
\end{center}
\caption{ (Color online)
  Plot of temperature profile and fourth moment of velocity
  for a strained $40\times 10$ lattice. The unstrained packing
  fraction was $\eta=0.85$ and the system was 
 strained to $\ex=0.0625$.
}
\label{LTE}
\end{figure}

We have also tested the assumptions of a linear temperature
profile and the assumption of local thermal equilibrium (LTE) that we
have used in our theory. In our simulations the local
temperature is defined from the local kinetic energy density,
{\emph{i.e.}} $\kb T = \la m \bu^2 /2  \ra$.  
Local thermal equilibrium requires a close to Gaussian distribution of
the local velocity with a width given by the same temperature. 
The assumption of LTE can thus be tested by
looking at higher moments of the velocity, evaluated locally. 
Thus we should have $ \la {\bu}^4 \ra = 8(\kb T/m)^2$.
From our simulation we find out $\la {\bu}^4(y)\ra$ and $\kb T(y)$
as functions of the distance $y$ from the cold to hot reservoir. The
plot in Fig.~\ref{LTE} shows that the temperature profile is
approximately linear and LTE is approximately valid.
We use our theory only in the solid phase and in this case there is
not much variation in the direction transverse to the direction of heat flow
($x$-direction).

\begin{figure}[t]
\begin{center}
\includegraphics[width=8.6cm]{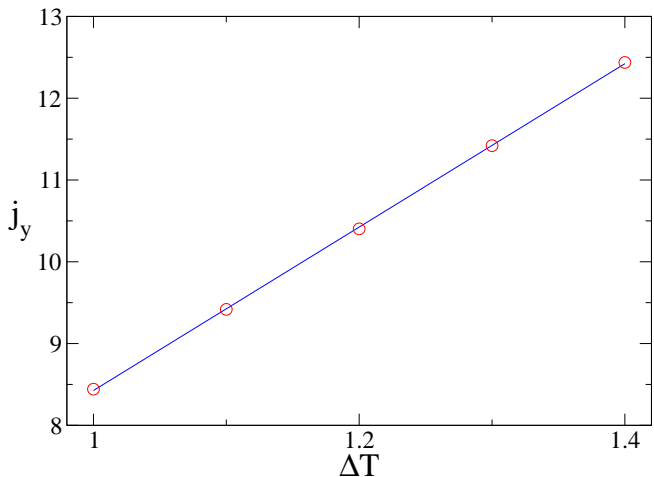}
\end{center}
\caption{ (Color online)
Plot of $j_y$ vs $\D T=T_2-T_1$ for a $40\times 10$ triangular
lattice at $\eta=0.85$. We see that the current increases linearly with
the applied temperature difference.  
}
\label{linresp}
\end{figure}

Finally we have verified that heat conduction in the small confined lattice
under small strains shows a linear response behaviour. This can be
seen in Fig.~\ref{linresp} where we plot $j_y$ vs  $\D T =
T_2-T_1$ for  a $40 \times 10$ triangular lattice at $\eta=0.85$.
Note that, as mentioned in the introduction, the bulk thermal
conductivity of a two-dimensional system is expected to be divergent
and the linear response behaviour observed here is only relevant for a
finite system and in a certain regime (solid under small strain).  

\section{Discussion}
\label{disc}
In this paper we have studied heat conduction in a two-dimensional solid 
formed from hard disks confined in a narrow structureless channel. The channel 
has a small width ($\sim 10$
particle layers) and is long ($\sim 100$ particles). Thus our system is in the
nanoscale regime.  
We have shown that structural changes that occur when this 
solid is strained can lead to sudden jumps in the heat current. 
From the system sizes that we have studied it is not possible to
conclude that these jumps will persist in the limit that the channel
length becomes infinite. However the finite size results are
interesting and relevant since real nano-sized solids {\em are} small.
We have also proposed a free volume theory type calculation of the heat
current. While being heuristic it gives correct order of magnitude
estimates and also reproduces qualitative trends in the current-strain graph.  
This simple approach should be useful in calculating the heat
conductivity of a hard sphere solid in the high density limit.

The property of large change of heat current  
could  be utilized to make a system perform as a mechanically controlled
switch of heat current. Similar results are also expected for the
electrical conductance and this is shown to be true at least
following one protocol of straining in Ref.\cite{my-econd}. 
From this point of view it seems worthwhile to perform similar
studies on  transport in confined nano-systems in three dimensions and
also with different interparticle interactions.  

\section{acknowledgment}
DC thanks S. Sengupta for useful discussions;
DC also thanks RRI, Bangalore for hospitality and CSIR, India for a fellowship.
Computation facility from DST grant No. SP/S2/M-20/2001 is gratefully 
acknowledged.
\vspace{0.35cm}
\appendix
\section{Calculation of $[y_c^2]_{fv}$ and $V_{fv}$}
Using free volume theory, as explained in the text, we get the
following expressions for 
$[y_c^2]_{fv}$ and $V_{fv}$:
\bea
[y_c^2]_{fv}&=&\f{1}{2}(\h+\psi+\phi) + \f{1}{4}\left(\sin 2\h + \cos 2\psi
             -\f{\sin 2\phi}{\h+\psi+\phi}\right) \nn\\
V_{fv} &=& -2(\h+\psi+\phi) - (\sin 2\h+ \sin 2\psi - \sin 2\phi) \nn\\
         &+& 4 a_y \left(a_x - \cos\phi\right)
\eea
where, 
$\h=\sin^{-1}(a_x/2)$, $\psi=\sin^{-1}(a_x/2 - \cos\phi)$ and
$\phi=\tan^{-1}(2a_y/a_x)-\cos^{-1}(l/2)$ with $l=\sqrt{(a_x/2)^2+a_y^2}$
 (all lengths are measured in units of $d$).
For strain along $x$-direction we have 
$a_x=a_x^0(1+\ex)$ and $a_y=a_y^0~ (= \sqrt3 a_x^0/2)$ while  
for strain along $y$-direction we have 
$a_y=a_y^0~(1+\ey)$ and $a_x=a_x^0$. 
From the above expressions we can obtain Taylor expansions of
$[y_c^2]_{fv}$ and $[\tau_c]_{fv} = c V_{fv}^{1/2}/T^{1/2}$, about the zero
strain value. These give the expressions for
$\a,~\be_1,~\be_2,~\g_1,~\g_2,~\g_3$  used in the text.

\end{document}